\documentclass[12pt,preprint,prl,aps,showpacs]{revtex4}

\usepackage{epsfig}

\begin{document}

\title{ $\rho^0$ meson production in the $pp \rightarrow pp\pi^+\pi^-$
  reaction at 3.67 GeV/c}

\author{
  F.~Balestra$^{1}$, Y.~Bedfer$^{2,a}$, R.~Bertini$^{1,2}$, L.C.~Bland$^3$,
  A.~Brenschede$^{4}$, F.~Brochard$^{2,b}$, M.P.~Bussa$^1$,
  Seonho~Choi$^{3,c}$, M.L.~Colantoni$^1$,  R.~Dressler$^{6,d}$,
  M.~Dzemidzic$^{3,h}$, J.-Cl.~Faivre$^{2,a}$,
  A.~Ferrero$^1$, L.~Ferrero$^1$, J.~Foryciarz$^{9,5,e}$, I.~Fr\"ohlich$^4$,
  V.~Frolov$^{7,f}$, R.~Garfagnini$^1$, A.~Grasso$^1$, S.~Heinz$^{1,2}$,
  W.W.~Jacobs$^3$, W.~K\"uhn$^4$, A.~Maggiora$^1$,
  M.~Maggiora$^1$, A.~Manara$^{1,2}$, D.~Panzieri$^8$, H.-W.~Pfaff$^{4,g}$,
  G.~Piragino$^1$, A.~Popov$^7$, J.~Ritman$^4$,
  P.~Salabura$^5$, V.~Tchalyshev$^7$, F.~Tosello$^1$, S.E.~Vigdor$^3$, and
  G.~Zosi$^1$}

\affiliation{
(DISTO Collaboration) \\
$^1$Dipartimento di Fisica ``A. Avogadro'' and INFN - Torino, Italy  \\
$^2$Laboratoire National Saturne, CEA Saclay, France\\
$^3$Indiana University Cyclotron Facility, Bloomington, Indiana, U.S.A.\\
$^4$II. Physikalisches Institut,  University of Gie\ss en, Germany \\
$^5$M. Smoluchowski Institute of Physics, Jagellonian University, Krak\'ow,
Poland \\
$^6$Forschungszentrum Rossendorf, Germany \\
$^7$JINR, Dubna, Russia \\
$^8$Universita' del Piemonte Orientale and INFN - Torino, Italy \\
$^9$H.Niewodniczanski Institute of Nuclear Physics, Krak\'ow, Poland \\
}

%\maketitle

\begin{abstract}
  Total and differential cross sections for the exclusive reaction
  $pp\rightarrow pp\rho$ observed via the $\pi^{+}\pi^{-}$ decay channel have
  been measured at $p_{beam}=3.67$~GeV/c.  The observed total meson production
  cross section is determined to be $(23.4\pm{0.8}\pm{8})\mu b$ and is
  significantly lower than typical cross sections used in model calculations
  for heavy ion collisions.  The differential cross sections measured indicate
  a strong anisotropy ($\sim cos^2\theta^{CM}_{\rho}$) in the $\rho^0$ meson production.
\end{abstract}


\maketitle

%\narrowtext

The study of the production of vector mesons (such as $\rho^0$) in
both hadronic and electromagnetic processes is considered an
excellent tool to investigate the properties of the hadrons both
inside the nuclear medium and in free space. These properties are
closely related to the QCD vacuum structure characterized by the
presence of a non-vanishing quark condensate breaking chiral
symmetry~\cite{Brown}. In QCD inspired models, in-medium $\rho^0$
mass modifications have been proposed as a signal for chiral
symmetry restoration in dense and/or hot baryonic matter
\cite{Brown}. Based on these calculations sizable changes in the
$\rho^0$ spectral function have been predicted for near-threshold
$\rho^0$ production in proton-, pion- and nucleus-nucleus
reactions ~\cite{li,cassing}. Furthermore modifications of the
in-medium $\rho^0$ spectral functions can also rise from $\rho-N$
coupling as indicated by hadronic model
calculations~\cite{rapp,rhospectral}. Such modifications could
have a significant impact on the role of the $\rho^0$ as one of
the mediators of the nuclear force at small distance, mainly in
the tensor part of the interaction~\cite{nak0,oset}.

A measurement of the $\rho$ production cross sections in $pp$ collisions
close to threshold is of particular importance for the understanding of the
$\rho-N$ coupling. Unfortunately, the large $\rho$ width $\Gamma=0.15$
GeV/$c^2$ and a small cross section hampered up to now successful $\rho^0$
identification for excess energies above threshold $\epsilon \le 1.1$ GeV
\cite{earlyrho}. In this letter, we report on the first measurement of the
exclusive total and differential cross sections of the $\rho^0$ production
in proton-proton reactions at $\epsilon = 0.33$ GeV, measured at the SATURNE
II facility at Saclay.

The proton beam of momentum 3.67 GeV/c was incident on a liquid
hydrogen target and charged particles were detected using the DISTO
spectrometer, which is described in detail elsewhere~\cite{NIM}. This
spectrometer consisted of a large dipole magnet (40 cm gap size, operating at
1.0 T$\cdot$m) which covered the target, located at the center, as well as two sets of scintillating
fiber hodoscopes. Outside the magnetic field, two sets of multi-wire
proportional chambers were mounted, along with segmented plastic
scintillator hodoscopes (SH) and water \v Cerenkov detectors (\v CD) for particle identification.

The acceptance of all detectors ($\simeq 2^\circ$ to $\simeq 48^\circ$
horizontally and $\simeq\pm 15^\circ$ vertically) on both sides of the beam,
allowed coincident detection of four charged particles with sizable
efficiency. The multi-particle trigger, which was based on the multiplicity
of SH elements and the scintillating fibers, selected events with at least
three charged tracks in the final state. With this trigger many exclusive
channels, e.g. $ppK^+K^-$, $pp\pi^+\pi^-\pi^0$ ~\cite{JIM1,JIM2,JIM3} and
$pK\Lambda$, $pK\Sigma$~\cite{DNN} were simultaneously measured.

The exclusive $\rho^0$ meson production ($pp\rightarrow pp\rho^0$)
was measured via its dominant ($\sim 100\%$) $\pi^{+}\pi^{-}$
decay channel. Proton and pion identification was achieved by
means of conditions defined on two dimensional distributions of
particle momentum versus \v Cerenkov light and energy loss
measurements in the \v CD and the SH, respectively. It was found
that less than $10\%$ of the pions could be misidentified protons.
The selection of the $pp\rightarrow pp\pi^{+}\pi^{-}$ reaction was
based on two kinematical conditions that were checked with Monte
Carlo simulations: (a) the four particle missing mass has to be
zero $(M^{miss}_{4p})^2= 0\pm 0.015$ GeV$^2/c^4$ and (b) the
proton-proton missing mass has to be equal to the $\pi^+\pi^-$
invariant mass $(M^{miss}_{pp})^2=(M^{inv}_{\pi^+\pi^-})^2 \pm
0.2$ GeV$^2/c^4$. The residual background below the
$(M^{miss}_{pp})^2$ bump amounts only to about $2\%$ of the total
yield and can mainly be attributed to events from the
$pp\pi^+\pi^-\pi^0$ reaction.

The acceptance of the DISTO spectrometer for the
$pp\pi^{+}\pi^{-}$ channel was determined using Monte Carlo
simulations, which after digitization, were processed through the
same analysis chain as the measured data.  The production of a
meson $X$ with the given mass $M_X$, in the exclusive reaction
$pp\rightarrow ppX \rightarrow pp\pi^+\pi^-$, was assumed for the
final state particles generation. Eight linearly independent
degrees of freedom are necessary to fully describe a four body
final state. We have chosen the following variables:
$(M^{inv}_{p_1X})^2,(M^{inv}_{p_2X})^2$ and three Euler angles
$\theta_X^{CM},\phi_X^{CM},\psi_{pp}^{CM}$ in the center of mass
frame (CM) that characterize the $ppX$ system, and
$M_{X},\theta_{\pi}^X,\phi_{\pi}^X$ which describe the
$X\rightarrow \pi^+\pi^-$ decay ~\cite{JIM3}.  The spectrometer
acceptance has been determined as a four dimensional function of
the variables $(M^{inv}_{p_1X})^2, (M^{inv}_{p_2X})^2,
\theta_X^{CM}$, and $M_{X}$, assuming isotropic distributions in
the remaining four variables. Symmetry reasons ensure that the
distribution of $\phi_X^{CM}$ must be isotropic, and the isotropic
assumption for the remaining three variables was corroborated by
the measured data.

  The simulations indicate a very low acceptance of the apparatus for
the $X$ produced in the backward hemisphere in the CM frame. However, since
the initial system consists of two identical particles a reflection symmetry
about $\theta^{CM}=90^{\circ}$ exists, thus the backward data are redundant
for the total cross section determination. In the forward hemisphere the
detector acceptance was found to be non-zero in every bin over the full
kinematically allowed region, thus eliminating the need for any model
dependent extrapolations of the cross section into unmeasured regions of
phase space. As a result, the acceptance corrections for the DISTO
spectrometer are independent of the actual final state distributions, except
for residual effects related to the size of the kinematic bins and the
detector resolution, which have been accounted for in the estimate of the
systematic error. Finally, the data presented below have been corrected on
an event-by-event basis, via a weighting factor obtained from the acceptance
function for the appropriate kinematic bin.

Compared to the other mesons identified in our experiment
\cite{JIM1,JIM2,JIM3} the extraction of the $\rho^0$ yield from the non
resonant background is much more difficult because of its larger width.
Therefore, different selection criteria had to be applied. For each of them
the effect on the acceptance for $\rho^0$ and background events has been
evaluated using the Monte Carlo simulation mentioned above. A first
selection was applied on the three body $p\pi^+\pi^-$ invariant mass
$M^{inv}_{p\pi^+\pi^-}>1.6$ GeV/$c^2$, for both $p\pi^+\pi^-$ combinations.
It was found that this rejects $35\%$ of the $pp\pi^+\pi^-$ events, whereas
the simulations indicate that only less than 4\% of $\rho^0$ events are lost
by this restriction.

\begin{figure}

 \begin{center}

  \mbox{\epsfig{figure={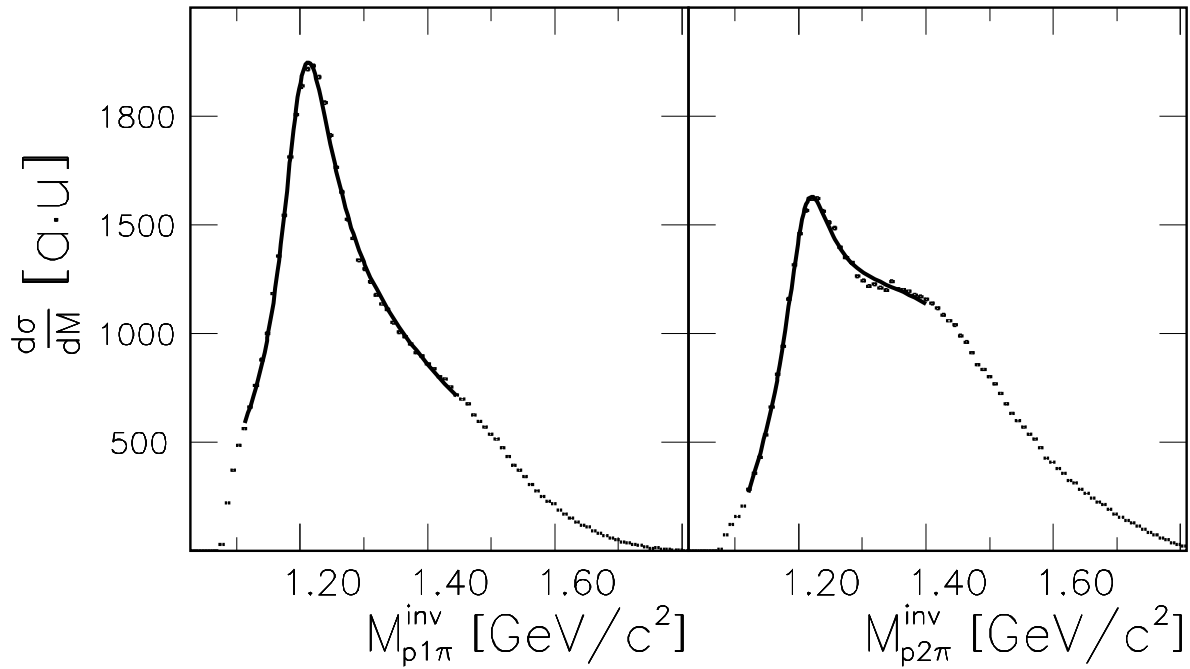},bbllx=1,bblly=17,bburx=384,bbury=256,width=0.99\linewidth}}

 \end{center}

 \caption{ Proton-pion invariant mass distributions $M^{inv}_{p\pi}$ selected
   according to the absolute value of the four-momentum transfer t:
   $M^{inv}_{p_1\pi^+}$ with the lower $\mid t\mid$ value (left frame) and
   $M^{inv}_{p_2\pi^+}$ with the larger $\mid t \mid$ (right frame).
   The curves are fits to the data as explained in the
   text.  }

 \label{prl1}

\end{figure}

A further investigation of the remaining events shows that, as
observed at higher energies~\cite{earlyrho}, most two pion events
originate from the $pp\rightarrow \Delta^{++} p\pi^-$ channel.
Figure~\ref{prl1} shows the acceptance corrected $p\pi^{+}$
invariant mass distributions for events fulfilling the
$M^{inv}_{p\pi^+\pi^-}>1.6$ GeV/c$^2$  condition discussed above.
The $M^{inv}_{p_1\pi^+}$ and $M^{inv}_{p_2\pi^+}$ have been
obtained selecting the events according to the absolute value of
the four momentum transfer $\mid t\mid$, from the incoming beam
proton to the outgoing $p\pi^+$ system.  $M^{inv}_{p_1\pi^+}$ and
$M^{inv}_{p_2\pi^+}$ denote the invariant mass of the p$\pi^+$
pair with smaller and larger momentum transfer, respectively.  The
prominent $\Delta^{++}$ peak at $M^{inv}_{p\pi^+}=1.231\pm0.008$
GeV/c$^2$ is visible in both distributions. A fit to the data
using a relativistic Breit-Wigner distribution~\cite{cassing} of
width $\Gamma=0.12$ GeV/c$^2$ added onto a third order polynomial
background shape indicates that $77\%$ of all $pp\pi^+\pi^-$
events involve $\Delta^{++}$ production.

\begin{figure}[ttt]

 \begin{center}

  \mbox{\epsfig{figure={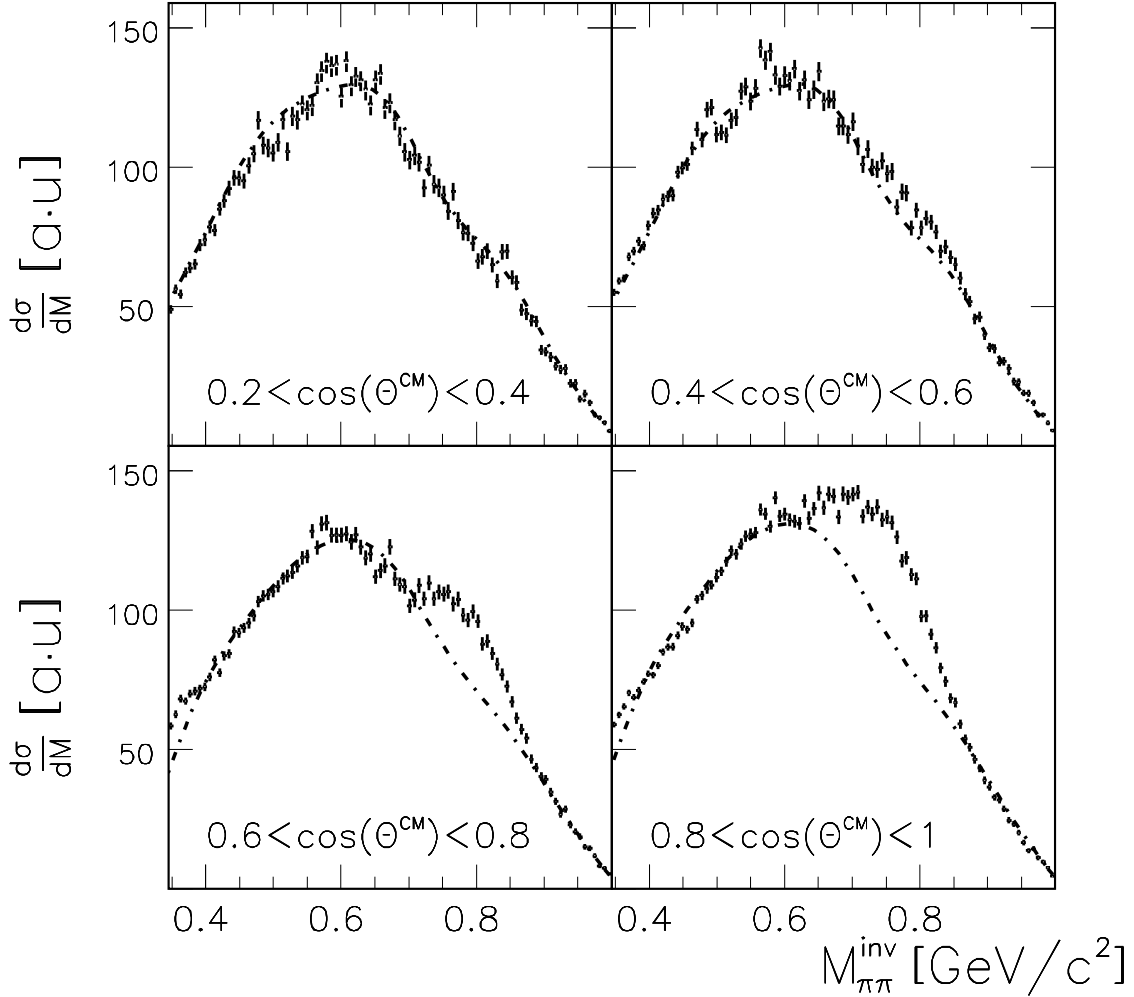},bbllx=1,bblly=13,bburx=414,bbury=390,width=0.95\linewidth }}

 \end{center}

 \caption{ $\pi^+\pi^-$ invariant mass distributions
          obtained in four different CM angular bins as indicated on
          the figure. The dashed line represents the smoothed invariant mass
          distribution for the $0<cos(\theta^{CM})<0.2$ normalized to the
          maximum.  }

 \label{prl2}

\end{figure}

In order to reduce the background related to $\Delta^{++}$
production, it was required that the invariant mass for both
$p\pi^+$ combinations be $1.15$ GeV/$c^2<M^{inv}_{p\pi^+}<1.28$
GeV/c$^2$ and that the three body invariant mass
$M^{inv}_{p_1\pi^+\pi^-}<1.74$ GeV/c$^2$. With such conditions
$50\%$ of the $\pi^+\pi^-$ pairs were rejected whereas the
simulation showed that only $10\%$ of those from $\rho^0$ decay
are lost. A similar selection on $p \pi^-$ invariant mass was
judged unnecessary, since events involving $\Delta^0$ production
are much less abundant than those with $\Delta^{++}$ and therefore
will not enhance the $\rho^0$ signal to background ratio.

For the events surviving all the criteria described above,
figure~\ref{prl2} presents the final, acceptance corrected
$M^{inv}_{\pi^+\pi^-}$ distributions for four equal bins in
$cos(\theta^{CM})$ between $cos(\theta^{CM})=0.2$ and
$cos(\theta^{CM})=1.0$, where $\theta^{CM}$ is the CM polar
emission angle of the di-pion system with respect to the beam
axis.  The dashed line shows the smoothed $M^{inv}_{\pi^+\pi^-}$
distribution for $0<cos(\theta^{CM})<0.2$ normalized to the
maximum, which reproduces well the general shape of the
$\pi^+\pi^-$ non resonant spectra. It should be noted that this
yield is almost independent of $cos(\theta^{CM})$ indicating a
nearly isotropic $\pi^+\pi^-$ non-resonant production. A structure
near the $\rho^0$ mass $M_{\rho}=0.77$ GeV/$c^2$, is clearly
visible.

\begin{figure}[ttt]

 \begin{center}

  \mbox{\epsfig{figure={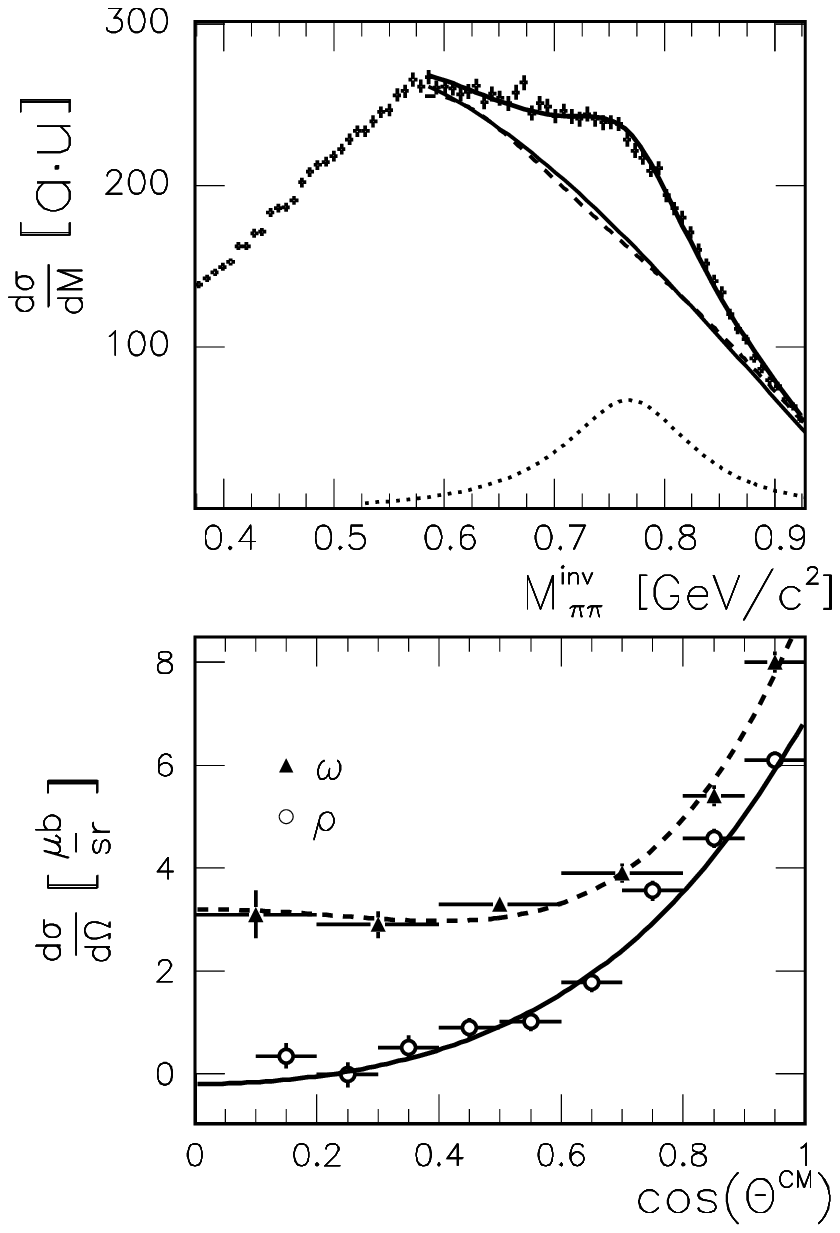},bbllx=1,bblly=24,bburx=267,bbury=405,width=0.9\linewidth }}

 \end{center}

 \caption{Upper frame: $M_{\pi^+\pi^-}$ distribution measured for
          $cos(\theta^{CM})>0.5$. The solid line shows the fit using a
          relativistic Breit-Wigner function added onto a third
          order polynomial for the background, as described in the
          text. The dotted line shows the meson line shape and the
          dashed line presents the $M^{inv}_{\pi^+\pi^-}$ distribution obtained
          for $cos(\theta^{CM})<0.4$ normalized to the same maximum.  Lower
          frame: The $\rho^0$ (circles) and $\omega$ (triangles) differential cross sections as a function of the
          meson CM emission polar angle. The solid and dashed lines show fits using the
          first three even Legendre polynomials.  }

 \label{prl3}

\end{figure}

The $\rho^0$ yield has been extracted from a fit to the $M^{inv}_{\pi^+\pi^-}$
distributions using a function $\sigma(m)= BW(m,\Gamma)\cdot V_{ps}(m)$
describing the $\rho^0$ line shape and a third order polynomial background.
The $ BW(m, \Gamma $) is a relativistic Breit-Wigner function with a mass
dependent pion decay width of the meson given by~\cite{cassing}:
$\Gamma=\Gamma(m_{R})\cdot(\frac{k_\pi(m)}{k_\pi(m_R)})^{3}\cdot\frac{m_R}{m}$,
where $\Gamma(m_R)=0.15$ GeV/c$^2$ is the full $\rho^0$ width at its pole
position $m_R=M_{\rho}$ and $k_\pi(m)$, $k_\pi(m_R)$ are pion three-momenta in
the rest frame of the resonance with mass $m$ and $m_R$, respectively.
$V_{ps}(m)$ stands for the phase space available for the direct (no
intermediate state) production of the meson with a given mass $m$.  We have
calculated $V_{ps}(m)$ by numerical integration of the simulated $pp \rho$
Dalitz distribution including the $\Delta^{++}$ suppression cuts discussed
above.  As an example, figure~\ref{prl3} shows the $M^{inv}_{\pi^+\pi^-}$
distribution for $cos(\theta^{CM})\geq 0.5$ containing most of the $\rho^0$
yield. The initial background parameters have been obtained by fitting the
$M^{inv}_{\pi^+\pi^-}$ distribution for $cos(\theta^{CM})\leq 0.4$ and
$M^{inv}_{\pi^+\pi^-}\geq 0.55$ GeV/c$^2$ normalized to the same maximum
(dashed line). The best fit ($\chi^2/\nu=1.2$) has been obtained with a fixed
$\Gamma(m_R)=0.15$ and $M_R=0.772\pm 0.003$ GeV/c$^2$ (dotted) and background
(solid line) which differs only very little from its initial shape.

In order to extract the angular distribution for the $\rho^0$
production and to extract the total production yield, a finer
angular bin size was chosen and the fit procedure outlined above
has been applied. The total $\rho^0$ yield has been obtained as a
sum of yields in the individual angular bins. The total cross
section of $\sigma=(23.4\pm 0.8\pm 8)\mu b$ with statistical and
systematic errors, respectively, was determined by normalizing the
acceptance corrected $\rho^0$ yield to the simultaneously measured
exclusive $\eta$ yield of known cross section $\sigma_{\eta}=135\pm35$ $\mu$b.
The $\eta$ yield
was obtained using separate acceptance correction matrices but
calculated with the same method \cite{JIM3}. This procedure to
determine the absolute normalization was chosen because it reduced
the large systematic uncertainty associated with the determination
of the absolute beam intensity and trigger efficiency. It was
checked by detailed Monte Carlo \hyphenation{si-mulations}
simulations that particle identification efficiencies and yield
reductions due to the analysis cuts applied for the selections of
various reactions are understood within $15\%$~\cite{JIM3}.
Finally, the method was verified by comparing the exclusive cross
section for $\omega$ production measured in our experiment
\cite{JIM3} with other experimental results obtained in a similar
energy range. An interpolation between existing data points give
$\sigma_{\omega}=45\pm 7\;\mu b$ ~\cite{cassib} in good agreement
with our result $\sigma_{\omega}=50\pm 3^{+18}_{-16} \mu b$.

The systematic uncertainty for the $\rho^0$ cross section contains
two almost equal contributions: (a) errors from absolute
normalization and (b) uncertainties related to the background
subtraction, acceptance corrections, tracking efficiency and
analysis cuts.  In particular a possible bias in the analysis due
to the $\Delta^{++}$ suppression has been investigated by
comparing the $\rho^0$ intensity obtained with and without the
suppression. The observed yield reduction of $\sim 10\%$ is
reproduced by the Monte Carlo simulations mentioned above, was
corrected for and completely included in the systematic errors.

Polar angle differential cross sections have been
proposed~\cite{nak} as a sensitive observable to extract the
strength of nucleonic and mesonic currents responsible in meson
exchange models for the vector meson production near the
threshold. Therefore, it is particularly interesting for various
reaction models to compare the differential cross sections for the
production of $\rho^0$ and $\omega$ as measured in our experiment
~\cite{JIM3}. The lower part of Figure~\ref{prl3} shows the
$\rho^0$ (circles) and $\omega$ (triangles) production cross
section as a function of $cos\theta^{CM}$, the polar angle in the
CM system of the meson. The vertical error bars reflect the
statistical errors, only. The distributions have been fit with the
sum of the first three Legendre Polynomials (solid line) $P_i$
with the result:
$\frac{d\sigma^{\rho}}{d\Omega}\propto(2.\pm0.2)P_0 +
(4.4\pm0.2)P_2 +(0.8\pm0.2)P_4$ and
$\frac{d\sigma^{\omega}}{d\Omega}\propto(4\pm0.1)P_0 +
(3.1\pm0.2)P_2 +(2.0\pm0.2)P_4$ for the $\rho^0$ and $\omega$,
respectively.

The above results indicate contribution of partial waves up to L=2
in the production of both mesons. Clearly, the angular
distribution for $\rho^0$ production exhibits stronger forward
peaking, suggesting a dominance of the nucleonic current
characterized by a dipole $\sim cos^2\theta^{CM}$ dependence
\cite{nak}. The angular distributions for $\omega$ production at
this beam energy and at the smaller excess energy $\epsilon=0.173$
GeV \cite{TOF} show strong nucleonic contribution too, but also an
additional isotropic component signaling the importance of the
mesonic $\pi\rho\rightarrow\omega$ fusion \cite{nak}. It is also
interesting to note, that the observed ratio of the total cross
sections for $\rho^0$ and $\omega$ production
$R=\sigma(\rho^0)/\sigma(\omega)=0.46\pm 0.12$ is very similar to
the ratio obtained at higher energy ($\epsilon=1.14$ GeV)
$R=0.48\pm 0.13$ ~\cite{earlyrho} and seems to indicate similar
scaling of the cross section with $\sqrt{s}$ for both mesons.

In conclusion, the production of the $\rho^0$ meson has been studied in the
$pp$ reaction at $p_{beam} = 3.67$ GeV/c.  A total $\rho^0$ cross section of
$\sigma=(23.4\pm 0.8\pm 8)\mu b$ has been determined.  It is significantly
lower than typical cross sections used by the hadronic models ($\sim 100$
$\mu b$) for meson production in pA and AA collisions
\cite{resmodel,cassing}. Furthermore, the differential cross sections
exhibit a prominent $\frac{d\sigma^{\rho}}{d\Omega}\propto cos^2\theta^{CM}$
dependence that might signal the dominance of nucleonic current in
$\rho^0,\omega$ meson production.

We thank the SATURNE II staff for delivering an excellent proton beam and
for assisting this experimental program.

This work has been supported by: CNRS-IN2P3, CEA-DSM, NSF,
INFN, KBN (5P03B14020) and GSI.

% bibliography%

\noindent

Now at: $^a$ DAPNIA/SPhN, CEA Saclay, France. \\
$^b$ LPHNHE, Ecole Polyt. 91128 Palaiseau, France.\\
$^c$ Temple Univ., Philadelphia, Pennsylvania, USA \\
$^d$ Paul Scherrer Institut, Villigen, CH-5232.\\
%$^e$ IU School of Medicine, Indianapolis, Indiana, USA.\\
$^e$ Motorola Polska Software Center, Krak\'ow, Poland.\\
$^f$ Dip. di Fisica ``A. Avogadro'' and INFN-Torino, Italy.\\
$^g$ Arthur Andersen, Eschborn, Germany.\\
$^h$ IU School of Medicine, Indianopolis,USA\\

\end{document}